\documentclass{emulateapj}
\usepackage{natbib}
\usepackage{amsmath}
\usepackage{subfigure}
\usepackage{soul,color}
\newcommand{\Mdot}{\dot{M}}
\newcommand{\OmK}{\Omega_{\rm K}}

\newcommand{\pfrac}[2]{\left(\frac{#1}{#2}\right)}

\newcommand{\rmag}{r_{\rm mag}}

\newcommand{\Rgas}{R_{\rm gas}}

\newcommand{\yr}{{\rm yr}}

\shorttitle{Stalling Type I Migration}
\shortauthors{Kretke \& Lin}

\begin{document}

\title{The Importance of Disk Structure in Stalling Type I Migration}
\author{Katherine A. Kretke$^{1,2}$, D. N. C. Lin$^{2,3}$}
\affil{$^1$Southwest Research Institute, Boulder, USA}
\affil{$^2$Department of Astronomy and Astrophysics, University
of California, Santa Cruz, USA}
\affil{$^3$Kavli Institute of Astronomy and Astrophysics, Peking
University, Beijing, China}
\email{kretke@boulder.swri.edu}

\begin{abstract}
As planets form they tidally interact with their natal disks.  
Though the tidal perturbation induced by Earth and super-Earth mass planets is generally too weak to significantly modify the structure of the disk,
the interaction is potentially strong enough to cause the planets to undergo rapid type I migration. 
This physical process may provide a source of short-period super-Earths, though it may also pose a challenge to the emergence and retention of cores on long-period orbits with sufficient mass to evolve into gas giants.
Previous numerical simulations have shown that the type I migration rate sensitively depends upon the circumstellar disk's properties, particularly the temperature and surface density gradients.
Here, we derive these structure parameters for 1) a self-consistent 
viscous-disk model based on a constant $\alpha$ prescription, 2) an 
irradiated disk model that takes into account heating due to the 
absorption of stellar photons, and 3) a layered-accretion disk model with 
variable $\alpha$ parameter.  We show that in the inner viscously-heated 
regions of typical protostellar disks, the horseshoe and 
corotation torques of super-Earths can exceed their differential Lindblad 
torque and cause them to undergo outward migration. However, the temperature profile due to passive stellar irradiation causes type I migration to be inwards throughout much of the disk. 
For disks in which there is outwards migration, we show that location and the mass range of the ``planet traps'' depends on some uncertain assumptions adopted for these disk models. Competing physical effects may lead to dispersion in super-Earths' mass-period distribution.
\end{abstract}

\section{Introduction}
\label{sec:introduction}
New high-precision radial velocity and transit surveys allow observers 
to detect low-mass and modest-size planets, particularly when those planets 
with short-period orbits.  Evidence is emerging that these close-in 
super-Earth mass planets are significantly more ubiquitous than their 
higher mass cousins, the hot Jupiters \citep{Mayor.etal.2009a, 
Mayor.etal.2011, Howard.etal.2010}.

Most theoretical models of hot Jupiter formation assume that planets 
form far from their parent stars and migrate inwards, either by 
planet-planet scattering \citep{Rasio.Ford.1996,Weidenschilling.Marzari.1996,
Lin.Ida.1997,Murray.etal.1998}, secular interactions such as the Kozai 
resonance \citep{Wu.Murray.2003}, secular chaos \citep{Nagasawa.etal.2008,
Marchi.etal.2009, Wu.Lithwick.2011}, or through tidal interactions with 
a gaseous circumstellar disk \citep{Lin.etal.1996}. This migration is 
required for hot Jupiters because it is difficult for them to form {\it in situ}.
The gas disk, which must be present to supply material to these planets, damps the eccentricity of planetesimals 
and embryos. This means that the growth of a protoplanetary embryo, on a circular orbit, 
stalls once it acquires an isolation mass ($M_{\rm iso}$), i.e. it has 
 captured all solids within a feeding zone centered around its orbital radius
$r$. The width of the feeding zone is approximately $\sim 10 r_{\rm Hill}$ where the 
Hill's radius for a planet of mass $M_p$ around a star of mass $M_*$ is 
$r_{\rm Hill} \equiv r[M_p/(3M_*)]^{1/3}$ \citep{Kokubo.Ida.2002}.
This leads to an isolation mass in a disk with the local surface density of dust and planetesimals ($\Sigma_d$) is $M_{\rm iso} \sim (\Sigma_d a^2/M_\ast)^{3/2} M_\ast$.

If we adopt a value for $\Sigma_d$ inferred from the empirical minimum 
mass nebula model \citep[MMN,][see below]{Hayashi.1981}, $M_{\rm iso}$ 
would be a fraction of Martian mass at 1AU and a few times Earth mass 
($M_\oplus$) at 5AU \citep{Ida.Lin.2004}.  This maximum mass increases with 
the distance from the central star as there are more solids 
available at large $r$ \citep{Lissauer.1987}. In the conventional 
core-accretion scenario, efficient gas accretion onto gas giants needs 
to be preceded by the emergence of cores with masses in excess of 
$\sim 10 M_\oplus$\citep{Pollack.etal.1996}. Progenitor cores with 
such large masses can only be assembled beyond the snow line in an MMN.

The same argument also suggests that it is difficult to form 
the rich population of known close-in (within $\sim 0.1$AU) 
super-Earths (a few to 15 $M_\oplus$) {\it in situ}. 
In order for there to be an adequate amount of building 
block material to assemble these planets the magnitude of $\Sigma_d$ must be more than an order of magnitude larger than that of the MMN.
If this necessary condition
is marginally satisfied then, in the presence of gas, instead of producing a single larger embryo, many embryos with much smaller isolation mass would emerge.
Under these conditions some rocky Earth analogs or super-Earths could have formed {\it in situ} through collisional mergers after the gas disperses (as it can no longer damp embryos' mutually excited eccentricities).
But giant impacts could still cause a substantial loss of volatiles in post-gas evolution,  so this method may not lead to the emergence of the known low-density super-Earths.

In order for the magnitude of $M_{\rm iso}$ to reach a few $M_\oplus$ at 0.1 AU, 
the local $\Sigma_d$ must be increased by two orders of magnitude over the MMN. 
In disks with a solar composition, the corresponding gas 
surface density ($\Sigma$) would be sufficiently large that gravitational 
instability would induce a rapid redistribution of mass and angular momentum \citep{Durisen.etal.2007, Gammie.2001, Boley.etal.2006}.  
Protoplanetary embryo formation in such disk regions may be challenged because the disk temperature would be above the condensation temperature of most refractory grains and the $\Sigma$ would decline substantially before the planetesimals can grow significantly.
In gravitationally stable disks with a $\Sigma$ distribution comparable
to that of the MMN, the dust to gas ratio would need to be enhanced 
by more than an order of magnitude relative to its solar value 
\citep{Ida.Lin.2005}.
This enhancement can be induced by planetesimal drift \citep[e.g.][]{Kretke.etal.2009,Hansen.Murray.2011} or by embryo migration. 

Planetary embryos migrate due to their interaction with gas in their natal
disks.  Their tidal perturbation excites density waves in their Lindblad 
and co-rotation (or horseshoe) resonance regions.  These waves carry
angular momentum flux which is determined by the surface density ($\Sigma$)
and temperature ($T$) of the disk gas at the excitation location 
\citep{Goldreich.Tremaine.1980}. 
These waves dissipate and deposit angular momentum
to the disk as they propagate away from the resonances 
\citep{Papaloizou.Lin.1984}.  
The disk responds with an evolving $\Sigma$ and $T$
distribution which introduces a feedback effect 
\citep{Takeuchi.etal.1996}.  
In order to calculate the full non-linear disk response, it is necessary 
to perform numerical hydrodynamic simulations of planets embedded 
in two or three dimensional disks.  
And mutual interactions between multiple planets embedded in a disk can lead to a wide varieties of diverse behaviors including outwards migration and migration stalling \citep[e.g.][]{Masset.Snellgrove.2001,Podlewska-Gaca.etal.2012}.
Nevertheless, if we restrict our investigation to planets with
modest masses, the disk structure is not strongly perturbed and
their net linear Lindblad and co-rotation torque 
can be calculated by summing up the contribution 
of each distinctive resonance. In disks with a power-law $\Sigma$ and 
$T$ profiles it is possible to describe the net torque on 
the planets as a linear combination of the radial $\Sigma$ and $T$ 
gradient \citep{Tanaka.etal.2002,Paardekooper.etal.2010}. 

In this paper, we are primarily interested in super Earths' type I 
migration.  The mass of these planets is generally too small to 
significantly perturb the structure of the disk.  In such limit, 
super Earths' migration rate can be deduced from the structure of 
unperturbed disk. Here, we apply various disk models to some recently
obtained prescriptions for type I migration \citep[][hereafer PBK10]
{Paardekooper.etal.2010a} and determine the direction
and rate of migration for planets with a range of masses.  

The simplest empirical prescription for the disk structure is the  
MMN model which is based the assumption that all planets in the 
solar system attained their mass {\it in situ} and they accreted 
all the heavy elements in their neighborhood \citep{Weidenschilling.1977a,Hayashi.1981}.  
Through the augmentation of volatile elements for a solar elemental 
distribution, we can infer a $\Sigma \propto r^{-3/2}$ distribution.  
With an additional assumption that protostellar disks are heated 
to their local equilibrium temperature, the disk temperature 
$T \propto r^{-1/2}$. The MMN is a reasonable fiducial model, albeit 
the {\it in situ} formation assumption is inconsistent with the 
current concept of planetary mobility.  

Alternatively, we can construct a set of dynamical models for the 
protostellar disks in which the $\Sigma$ and $T$ distribution are
determined with physical principles. 
Due to uncertainties in the mechanism of angular momentum in disk it is common to use the $\alpha$-viscosity model \citep{Shakura.Sunyaev.1973}  in which one assumes that the disk evolves due to a viscosity proportional to the sound speed ($c_s$) and disk scale height($h$),
\begin{equation}
\nu = \alpha c_s h.
\label{eq:nu}
\end{equation}
For the disk models we adopt here we first present results for disks with the common assumption that $\alpha$ is constant.
We then investigate disks in which we assume the magneto-rotational instability (MRI) excites turbulence and leads to angular momentum transfer in  regions with sufficiently high ionization fraction \citep{Hawley.etal.1995}.
In most planet forming parts of the disk (ranging from a fraction to a few 10 or AUs), 
MRI active regions are confined to the surface layers which are 
exposed to ionizing stellar photons and cosmic rays \citep{Gammie.1996}.  
In order to include this effect, we generalize the quasi steady state 
accretion disk model to allow for a variable $\alpha$ parameter.

There are a number of groups who have presented sophisticated structure 
models of MRI-active disks \citep{Sano.etal.2000,Terquem.2008,Kretke.Lin.2010}.
Due to the uncertainties on both MHD turbulence and dust grains'
size distribution in the disk surface layer, we choose to use a model 
with fewer assumptions to simplify this issue.  

A few studies have already begun investigating planet migration in 
MRI-active disks including the effects of a dead zone.  
\citet{Kuchner.Lecar.2002} and \citet{Masset.etal.2006a} investigated 
planet stalling at the inner edge of the dead zone.  
\citet{Matsumura.etal.2007,Matsumura.etal.2009} studied the growth 
and migration of Jovian planets in disks with dead zones. All of these 
studies indicate that planet migration is sensitive to the structure 
of the disk.  In this paper we present the evolution in simple disk models 
in order to discuss the generic disk profiles that can create ``planet 
traps'' and how these are likely to manifest themselves in physical disks.  

As we have indicated above, the rate and direction of 
type I migration is determined by the competing contribution from 
differential Lindblad and corotatation (or horseshoe) torque.  In
our computation of the condition for corotation torque saturation, 
we are mindful that the horseshoe region has a toroidal shape.  Its 
height normal to the disk plane is comparable to its width in radial 
direction. For planets with mass less than a few $M_\oplus$,
the entire horseshoe region may be embedded in the dead zone.
It is not clear whether there is any significant viscous flow across the
horseshoe zone to prevent saturation.  In these low-viscosity regions, 
embedded planets' long-term interaction may also modify the $\Sigma$ 
distribution, trigger secondary instabilities, and halt their migration
\citep{Balmforth.Korycansky.2001, Li.etal.2005, Li.etal.2009, Yu.etal.2010}.

This paper is organized as follows.
In \S \ref{sec:migration}, we briefly recapitulate relevant migration 
prescriptions.
In \S \ref{sec:model}, we construct three sets of disk models
which take into account surface irradiation and the presence of
the ``dead'' midplane between MRI active surface layers.
In \S \ref{sec:results}, we present the results on how planets 
will migrate in these models of static disks.
In \S\ref{sec:uncertainties}, we discuss potential complication
introduced by layer structure in the dead zones and feedback. 
In \S\ref{sec:pop}, we discuss how this work connects to the ongoing effort of population synthesis modeling.
In \S \ref{sec:summary} we summarize our results and discuss their 
implications.

\section{Migration Rates}\label{sec:migration}
The type I migration rate is sensitive to the local $\Sigma$ and $T$ 
as well as to the radial gradients of these quantities.  Following 
PBK10, we adopt the following general expression for the rate of 
type I migration,
\begin{equation}
	\frac{dr}{dt} = f(p,q,p_\nu,p_\chi) 
\frac{M_p}{M_*}\frac{\Sigma r^2}{M_*} \pfrac{r\OmK}{c_s}^2 r\OmK,
\end{equation}
where $M_p$ and $M_*$ are the planet and stellar mass, respectively, 
$\Sigma$, $T$, $c_s$, and $\OmK$ all refer to the those quantities at 
the location of the planet. The magnitude and sign of the dimensionless 
efficiency factor $f$ is a function of the dimensionless
logarithmic surface density and temperature gradients, $p(r) \equiv 
d\log\Sigma/d\log r$ and $q(r) \equiv d\log T/d\log r$, respectively.
Note that the PBK10 $\alpha$ and $\beta$ parameters correspond 
to $-p$ and $-q$, respectively, in this paper.

In weakly perturbed disks, $p$ and $q$ are intrinsic properties of 
the disk alone. The efficiency factor $f$ is also a function of the  
viscous and thermal saturation parameters 
\begin{equation}
	p_\nu \equiv \frac{2}{3} ({\rm Re}~x_s^3)^{1/2}
	\label{eq:p_nu}
\end{equation} 
and 
\begin{equation}
p_\chi \equiv (\OmK r^2 x_s^3 /2 \pi \chi)^{1/2}=3 p_\nu / 2\rm P_r^{1/2}
\label{eq:pchi}
\end{equation}
where ${\rm Re} \equiv \OmK r^2/2 \pi \nu$ is the Reynolds number, $x_s
\simeq (M_p r/M_\ast h)^{1/2}$ is the dimensionless width of the 
horseshoe orbit, $h=c_s\OmK^{-1}$ is the density scale height in the 
direction normal to the plane ({\it i.e.}~the thickness) of the disk,
and $P_r=\nu\chi^{-1}$ is the Prandtl number.
The magnitude of both $p_\nu$ and $p_\chi$ depend on the mass ratio
$M_p/M_\ast$ as well as disk parameters, such that 
\begin{equation}
p_\nu \propto\alpha^{-1/2}(M_p/M_*)^{3/4}(h/r)^{-7/4}
\label{eq:pnu}
\end{equation}
in a disk in which the viscosity is parameterized according to eq. \ref{eq:nu}.

Based on a 3D linear analysis of wave excitation in a locally isothermal 
disk, \citet[][hereafter TTW02]{Tanaka.etal.2002} determined the 
migration rate to be
\begin{equation}
	f_{TTW}(p,q) = 1.08(p+0.80q-2.52).
\end{equation}
In typical disks, differential Lindblad resonances generally lead to 
inward migration.  Under some conditions, contribution from co-rotation
resonances and horseshoe torques can lead to outward migration.
\citet[][hereafer PBCK10]{Paardekooper.etal.2010} investigated disks in 
which there are fully-unsaturated ({\it i.e.}~with a maximum efficiency) 
horseshoe torques.  From these upper torque limits, we can identify disk
regions where planets have a {\it potential} to undergo outward migration.
In a similar (to TTW02), locally isothermal limit, PBCK10 found that the 
combined contribution from linear Lindblad torques, linear-entropy related 
co-rotation torque, and non-linear vortensity related horseshoe drag 
(with a softening factor $b/h = 0.4 $) leads to
\begin{equation}
	f_{\rm iso}(p,q) = -1.7-1.8q + 2p.
\end{equation}

PBCK10 also considered the effects of disk thermal structure.  They found 
that in an inefficiently radiating disk with an adiabatic equation of 
state, a planet migrating under the fully unsaturated ({\it i.e.}~with full 
strength) non-linear horseshoe torques (with the adiabatic index 
$\gamma=1.4$) yields
\begin{equation}
	f_{\rm ad}(p,q) = -1.22-5.64q+4.66p.
\end{equation}
With these migration laws, it becomes simple to identify the necessary 
conditions for stalling migration.  The necessary criteria for outwards 
migration as a function of $p$ and $q$ are listed in table \ref{table:pq}. 
For comparison and illustration purposes, we also included in this table 
the criteria relevant to the radial drift of particles due to aerodynamic 
drag \citep{Weidenschilling.1977}.

The results in table \ref{table:pq} indicate that with an adiabatic 
equation of state, outward migration in disks with negative surface 
density gradients ($p<0$) is only possible if there is a strong 
negative temperature gradient ($q< -0.22$).  Furthermore, in order 
for the torques to reach these fully unsaturated values, angular 
momentum must be extracted from the horseshoe region at the optimal 
rates, {\it i.e.}~fast enough to avoid torque saturation, but slow 
enough to maintain a gradient across the horseshoe region. These
requirements determine a set of sufficient conditions for stalling 
type I migration.  

\begin{table}
\begin{tabular}{ll}
Reference & Criteria\\
\hline
Radial Drift & $p > 1.5 -0.5 q$ \\
TTW02 & $p > 2.52 -0.8q$  \\
PCBK10 Locally Isothermal & $p > 0.85-0.9q$ \\
PCBK Adiabatic & $p > 0.26+1.21q$\\
\end{tabular}
\caption{Conditions for Outward Migration}
\label{table:pq}
\end{table}

\begin{figure}
\includegraphics[width=\columnwidth]{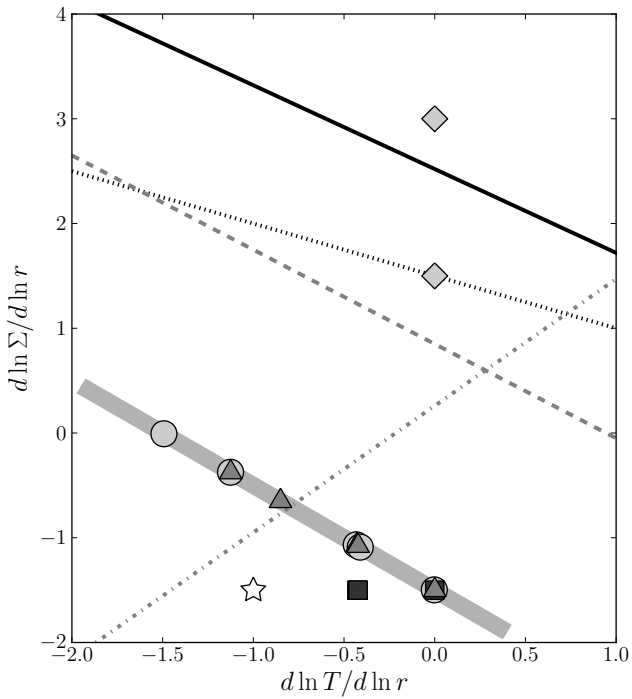}
\caption[A comparison of disk structure models and migration models]
{The surface density ($p$) and temperature ($q$) gradients for various 
disk models described in the text ({\it Star}-MMSN, {\it Squares}-CG97, 
{\it Circles}-Viscous Disk, {\it Triangles}-GL07. {\it Diamonds}-see 
\S\ref{sec:model_MRI}) compared to the criteria for zero-migration 
under the different prescriptions presented in table \ref{table:pq} 
({\it Solid} - TTW02, {\it dotted} - Aerodynamic Drag, {\it dashed} 
- Isothermal PBCK10, {\it dash-dotted} - Adiabatic PBCK10).  The 
shaded rectangle indicates the location of constant $\alpha$ models.}
\label{fig:intro}
\end{figure}

\section{Disk Models}\label{sec:model}
Here, we highlight migration rate's sensitive dependence on the 
disk structure. In the introduction, we have already indicated that 
the $\Sigma$ and $T$ distributions are determined by the 
not-fully-understood physics associated with the MRI turbulence 
and surface irradiation. In order to identify some robust features and assess
range of uncertainties, we adopt three different disk models including 1) 
a relatively simple self-consistent constant $\alpha$ model 
\citep{Shakura.Sunyaev.1973, Lin.1981} (see \S \ref{sec:model_vis}) and 2) a model for the irradiated outer regions of the disk (see \S \ref{sec:model_GL}). We also consider 3) a disk model which 
takes into account the influence of a dead zone (see \S\ref{sec:model_MRI}).  

Before describing these three models, we orient the reader by describing 
some standard disk models and features common to most disk models.
In figure \ref{fig:intro} we show where various disk models lie 
on the $p$-$q$ plane. The $\Sigma$ profile of the minimum 
mass nebula corresponds to $p=-1.5$, 
and on this plot we place a star to indicate a geometrically flat 
($q=-1$) MMN disk.  It is interesting to note that in a MMN profile, 
the vortensity gradient is zero and the horseshoe toques do not contribute
(see further discussion in \S\ref{sec:uncertainties}).

The observed spectral energy distributions (SED) of young stars indicate 
that at least the outer regions of most protostellar disks have  
a flaring structure. This feature means that $h/r$ increases with 
$r$, an indication that these passive disks have slowly declining 
temperature gradients ($q > -1$).  
\citep[][hereafter referred to as CG97]{Chiang.Goldreich.1997} construct a model to match the observed 
features.  In their model, there are three regions. Far from the star the 
disk is optically thin to both its own radiation and the stellar radiation, 
while in the inner region is optically think to both.  In between these 
two boundaries,  there is a marginally opaque region in which the disk 
is optically thick to the stellar photons, while optically thin to the 
longer wavelength disk photons (see \S\ref{sec:model_GL}).
The square symbols in Figure \ref{fig:intro}
indicate the temperature gradients of these three regions with 
an assumed MMN $\Sigma$ profile.  In this figure, the opaque and 
optically thin limits are virtually indistinguishable. 

\subsection{Steady state disk models with constant $\alpha$}

Instead of assuming a fixed $\Sigma$ profile (such as that inferred 
empirically with the MMN scenario), it is also possible to
deduce it from accretion disk models.  In an viscous axisymmetric 
Keplerian disk, the mass accretion rate through each annulus is given by
\begin{equation}
	\dot M = 3\pi\nu\Sigma = 3\pi\alpha c_s^2 \OmK^{-1}\Sigma.
\end{equation}
We assume short-period super-Earths were formed within a few AUs prior 
to their type I migration. In these inner disk regions, 
gas attains a quasi steady state, {\it i.e.}~its mass accretion rate
($\dot M$) is independent of disk radius \citep{Ruden.Lin.1986} albeit 
$\dot M$ does evolve over time.  In such a state, $\Sigma$ is inversely
proportional to the effective viscosity $\nu$.  

It is generally assume that turbulence provides an effective angular 
momentum transfer mechanism and the magnitude of viscosity can be 
approximated with an {\it ad hoc} $\alpha$ prescription in which 
$\nu = \alpha c_s H$ where $c_s\equiv (\Rgas T/\mu)^{1/2}$ ($\Rgas$ 
and $\mu$ are the gas constant and the mean molecular weight), 
$H=c_s/\OmK$ and $\OmK \equiv (G M_*/r^3)^{1/2}$ are the sound speed, 
density scale height and angular frequency of the disk gas
\citep{Shakura.Sunyaev.1973}. If the efficiency factor $\alpha$ 
is constant with radius, this steady state would lead to a simple 
relationship between the surface density gradient and temperature 
gradient, such that \linebreak $p=-q-3/2$.  The shaded rectangle in 
figure \ref{fig:intro} indicates the parameter space occupied by 
steady-state, constant $\alpha$ models.

In order to deduce a $\Sigma$ distribution, it is necessary to independently 
or concurrently determine a $T$ distribution. One possible approach is to 
assume an equilibrium $T \propto r^{-1/2}$ distribution as in the MMN model
and derive a $\Sigma \propto r^{-1}$ distribution \citep{Hartmann.etal.1998}.
Although this equilibrium $T$ distribution is adequate for the 
optically thin outer regions of the disk, it is not appropriate for the
inner disk regions which is mostly heated by viscous dissipation.
In standard accretion disk models, both $T$ and $\Sigma$ distribution
can be obtained self consistently under the assumption of thermal 
equilibrium, {\it i.e.}~viscous dissipation is balanced by local radiative 
flux \citep{Shakura.Sunyaev.1973, Lin.1981}.  

It is interesting to note that, although these standard disk models all 
lie in the lower region of this parameter space, the entire region 
indicated in this diagram satisfies the stability criteria: 
$d/dr[(r^2\Omega)^2] > 0$ (Rayleigh's criterion) and $d\rho/dr > d\rho/dz$.
Thus, the replacement of a constant $\alpha$ by a more general prescription
does not generally introduce dynamically unrealizable models.

\subsection{Description of viscously-heated disk model} \label{sec:model_vis}
In order to determine temperature of a viscously heated disk, we balance the radiative losses at the surface (of temperature $T_{\rm eff}$) with viscous heating and heating due to the background radiation, meaning
\begin{equation}
	2\sigma_{\rm SB} T_{\rm eff}^4 = \frac{9}{4} \Sigma \nu \OmK^2+2\sigma_{\rm SB}T_b^4
\end{equation}
where $\sigma_{\rm SB}$ is the Stefan-Boltzman constant and $T_b$ is the background temperature. 
The effective temperature  can be related to the midplane temperature via the effective optical depth $T_{\rm eff}^4 \tau_{\rm eff} = T^4$, where
\begin{equation}
	\tau_{\rm eff} \equiv \frac{3}{8}\tau + \frac{\sqrt{3}}{4} + \frac{1}{4\tau}
\end{equation}
and $\tau = \kappa \Sigma /2$ is the true optical depth \citep{Hubeny.1990}.
For all of the disk models we parameterize the opacity as
\begin{equation}
\kappa(T) = \kappa_0 T^\beta.
\label{eq:beta}
\end{equation}

\subsection{Description of GL07 model of irradiated disks} \label{sec:model_GL}
\begin{figure}
	\includegraphics[width=\columnwidth]{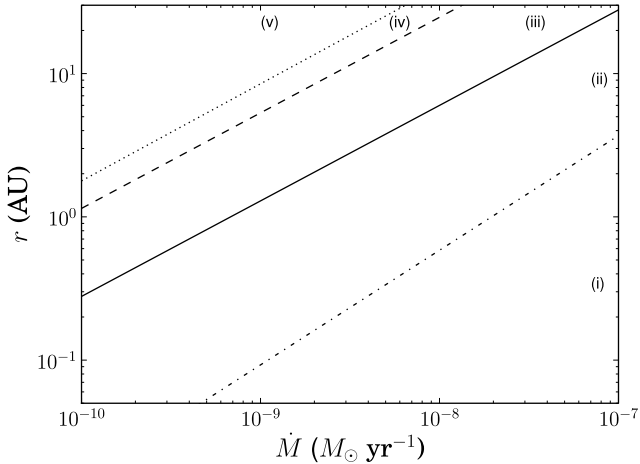}
	\caption[Evolution of transition radii in the GL07 disk]
	{Evolution of transition radii as a function of mass-accretion rate.  
	The transition between [i] viscously heated, [ii] strongly opaque, [iii] weakly opaque, [iv] marginally opaque, and [v] optically thin are marked with the {\it dashed-dot},  {\it solid}, {\it dashed}, and {\it dotted} lines.}
	\label{fig:GL07radii}
\end{figure}

In the determination of the $\Sigma$ and $T$ distribution for the 
intermediate and outer disk regions, contribution from stellar 
irradiation needs to be taken into account. In an attempt to introduce 
a more comprehensive prescription, \citet{Garaud.Lin.2007} constructed 
an analytic disk model (hereafter referred to as GL07) based on the 
passive disk model of CG97.  However, instead of assuming a MMN surface 
density profile, GL07 calculated the structure of a disk evolving with 
a constant $\alpha$-viscosity.  Additionally, this model includes both 
stellar irradiation and viscous heating.  This prescription results in 
five distinct regions.  The outer three regions, the optically thin 
[v], marginally opaque [iv], and weakly optically thick [iii] regions 
are similar to those in CG97.  In these regions the $T$ distribution, 
is independent of the $\Sigma$ profile, although a steady state 
solution can be preserved with a variable $\alpha$ prescription.

In the GL07 model, there are two regions close to the star , that do not 
occur in the CG97 passive-disk model,  a strongly-opaque region [ii] 
and viscously-heated region [i].  In region [ii], the optical depth to the 
disks own radiation is high, causing the disk to be non-isothermal both 
in the disk plane and in the direction normal to it.  In region 
[i], viscous dissipation dominates the thermal profile.
In these regions $q$ is a function of $p$.  In the inner disk regions
where accretion flow is in a quasi-steady state, both quantities
($p$ and $q$) are well defined. 

These five regions ([v] and [iii] are nearly indistinguishable) are 
indicated by triangles in figure \ref{fig:intro}.  The radial 
transitions between these regions evolve as a function of the 
mass accretion rate (see figure \ref{fig:GL07radii}).
It is interesting to note that the optically thick region of GL07 
lies remarkably close to the separatrix between outwards and 
inwards migration in figure \ref{fig:intro}.

\subsection{Description of an MRI-active Disk Model}\label{sec:model_MRI}
As we have indicated above, the $T$ distribution in the inner disk 
regions is determined by viscous dissipation and that in the outer disk
regions is determined by stellar irradiation.  The evolution of the
$\Sigma$ distribution everywhere in the disk is determined by 
the efficiency of angular momentum transfer, {\it i.e.}~the magnitude of
effective viscosity $\nu$, which we assume is mainly due to MRI driven MHD
turbulence.  In the innermost regions of protostellar disks where hot
Jupiters and super Earths eventually reside, the ionization fraction 
of the disk gas is modest and MHD turbulence is effective throughout 
the disk. But, beyond a fraction of an AU where most planets form and
start their migration, disks' midplane is generally too cool and neutral 
to be affected by magnetic fields.  However, their surface layers are 
exposed to the stellar UV photons and sufficiently ionized for 
MRI-driven MHD turbulence to prevail.  Wave propagation from the 
surface layer to the midplane can also lead to angular momentum, albeit
at much reduced rates\citep{Fleming.Stone.2003,Turner.etal.2007}.

Dynamical evolution of MHD turbulence in protostellar disks is beyond 
the scope of this paper.  In lieu of a detailed turbulent-disk model, 
we simply adopt a variable $\alpha$ in disk regions extend over the 
migration paths of hot Jupiters and super Earth. This {\it ad hoc} 
procedure is constructed in the same spirit as the original $\alpha$
prescription \citep{Shakura.Sunyaev.1973}, {\it i.e.}~for computational 
simplicity.  To do so, we introduce a parameter $\zeta(r)\equiv 
d\log\alpha/d\log r$ which varies with $r$ depending on the ionization
of the disk gas.  In the appendix, we present a simple modification 
to the GL07 model to allow for a variable $\alpha$.  

In locations where thermal ionization becomes sufficient to sustain 
MRI ($\sim 10^3$K) and at the sublimation front where dust destruction 
decreases the recombination rate, the magnitude of $\alpha$ increases 
rapidly with the disk temperature. If viscous dissipation is the dominant 
contributer to the temperature profile in these locations ({\it e.g.} 
in the regions [i] and [ii]), the disk would tend to evolve and attain 
a relatively shallow radial $T$ gradient.  In the limit that $q$ is 
negligible, $\zeta_{\rm V} = -9/2$ and $\zeta_{\rm SO}=-3$, leading 
to a $p_{\rm V}= 3$ and $p_{\rm SO}= 3/2$ for the viscous and strongly opaque regions, respectively.  In both cases, 
fully unsaturated horse shoe torque would induce a rapid outward migration.

\subsection{Inner Disk Truncation}\label{sec:rmag}
A generic feature of protostellar disks is the presence of a cavity
at their center which is cleared by the magnetosphere of their central
stars. The presence of this magnetospheric cavity quenches both
Lindblad and horse shoe torques and therefore provides a natural
post for hot Jupiters' type II migration to stall \citep{Lin.etal.1996}.
Here, we consider the possibility that the termination of super Earths'
type I migration may have also been strongly influenced by this structure.

As an additional feature in our disk models, we consider the possibility
of an inner boundary due to the disks' interaction with the stellar 
magnetosphere of their central stars.  The location of this magnetospheric 
truncation radius is assumed to occur at
\begin{equation}
	\rmag \simeq 0.5 \mu_*^{4/7} (2 G M_*)^{1/7} \Mdot^{-2/7},
	\label{eq:rmag}
\end{equation}
where $\mu_*$ is the stellar magnetic dipole moment \citep{Koenigl.1991}.

Many of the approximations integral to the disk model break down in this 
location.  The actual $\Sigma$ and $T$ profiles in this disk region need 
a detailed analysis of the process of mass loading onto the stellar magnetic 
field as well as magnetic diffusivity and dissipation near the disk inner 
boundary.  These studies are beyond the scope of this paper.  For the 
purpose of studying the destiny of super Earths, we impose 
a sharp positive surface density gradient near $\rmag$, where the surface density is reduced by the factor $f_{\rm mag}$.  In this paper we use
\begin{equation}
	f_{\rm mag} \equiv \exp\left(-\frac{1}{2}\pfrac{r-\rmag}{0.1 \rmag}^2\right),
\end{equation}
though we note that this is an arbitrary choice.
And as noted by \citep{Tsang.2011} the interaction at the inner edge of the disk are likely more complicated than the simple picture presented here.  As a planet approaches the inner edge of the disk, it is possible that reflections at the inner disk edge may cause the planets to have their migration halted at a few times the inner disk radius.  
But while the details are different, the unsaturated horse shoe torque near $\rmag$ generally induce a rapid outward type I migration, creating a planet trap.

\section{Type I Migration Rates deduced for Various Disk 
Models}\label{sec:results}
\subsection{The Migration Rate of a Planet in a Steady-State Disk}
In order to discuss the dominant and relevant physical effects, we 
illustrate the migration of a planet in a static ({\it i.e.}~a
steady-state) disk.  In figure \ref{fig:notrack_viscous}, we show 
the relative migration rates (as compared to the TTW02 model) for 
a disk which is heated entirely by viscous dissipation (see \S\ref{sec:model_vis}).  Such a 
structure is possible if the stellar irradiation is blocked by
self shadowing effects \citep{Dullemond.etal.2002}.  We 
adopt the following model parameters: $\dot M = 10^{-8} 
M_\sun \yr^{-1}$, $\alpha=10^{-3}$, $T_b=20$K, and $r_{\rm mag} = 0.15$ AU.  

From this standard viscous disk model, we find distinct regions 
where the disk structure can allow outwards migration.  Other than the 
inner boundary of the disk, these locations include the regions 
far from the sublimation temperatures, {\it i.e.}~zone 1 where 
the opacity is dominated by silicates ($\beta=1/2$ in eq \ref{eq:beta}) and zone 2 
where the opacity is dominated by ice grains ($\beta=2$).  
In the regions interior to zone 1 (where silicate grains partially 
sublimate), between zones 1 and 2 (where ice grains partially 
sublimate), and outside zone 2 (where the disk become optically
thin in the direction normal to the disk), fully unsaturated 
horse shoe torques cannot offset the differential Lindblad 
torque to induce planets to migrate inward.
.
\subsection{Potential location of planet traps}  
Fully unsaturated horse shoe torque has a tendency to induce 
planets formed in zones 1 and 2 to converge with those formed 
beyond these zones.  However, due to saturation effects, only 
planets with a limited range of masses can experience the full 
horse shoe torques.  For example, the dynamics of very low-mass 
planets is not affected by the horse shoe torques and they 
migrate inwards everywhere in the disk, albeit slowly, under 
the influence of differential Lindblad torque.  Horse shoe 
torques are also saturated for relatively massive planets 
(with masses at least an order of magnitude larger than that 
of Earth) and they undergo rapid inward migration everywhere
except near the disks' inner boundary.  Their inward migration 
would be stalled at $\rmag$ if the surface density fall off 
rapidly interior to the edge of the disks' inner cavity.

\subsection{Migration during planetary growth}
\begin{figure}
	\includegraphics[width=\columnwidth]{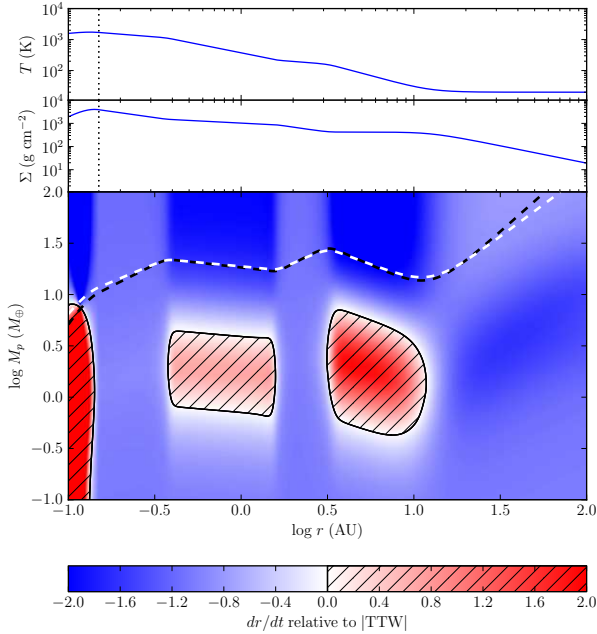}
	\caption{The temperature (top), surface density (middle), and 
migration rate relative to the magnitude of TTW02 as a function of radius and planetary mass in a  viscously heated disk (bottom).  We hatch the region in which migration is outwards for emphasis.
    The magnetospheric truncation radius is indicated by the dotted vertical line.
	The black dashed-line indicates when the disk interaction 
becomes non-linear (eq. \ref{eq:nonlinear}) and the white dashed-line 
indicates the planet mass which satisfies the gap opening criteria 
(eq. \ref{eq:gapopening}).  
In this example the two criteria are virtually identical, but this is not the case in general disks.}\label{fig:notrack_viscous}
\end{figure}

We now consider the implication of the results in figure 
\ref{fig:notrack_viscous} on the growth of planetesimals
and protoplanetary embryos.  For this model ({\it i.e.}~this
choice of $\dot M$ and $\alpha$), all planetesimals with mass larger
than that of Mars migrate inward at rates comparable to that
obtained by TTW02.  However, with adequate metallicity, the 
characteristic growth time scale for the modest-mass planetesimals 
\citep{Ida.Lin.2004} is shorter than their migration time scale
such that they can grow {\it in situ} into dynamically 
isolated protoplanetary embryos.

Embryos' isolation mass increases rapidly with $r$. Outside an 
AU or so, the it exceeds Mars' mass. Based on the results
in figure \ref{fig:notrack_viscous}, we find that the partially
saturated horse shoe torque induced on these modest-mass embryos
is adequate to cause an outward migration towards the outer 
boundaries of zones 1 and 2.

The migration of these embryos would be stalled at these 
planet-trapping boundaries.  In this set of models, there is an 
arbitrary growth barrier.  Once a core is stalled at a barrier,
it would accrete all of the material in its vicinity until 
it has attained a new isolation mass ($M_{\rm iso}$ is modified 
by the enhanced $\Sigma_d$ due to the accumulation of 
planetesimals and embryos).  In a disk with a single core there is no 
physical mechanism for it to grow beyond this size.

In this model where the disk is entirely heated by viscous 
dissipation, it is possible to trap super Earth embryos beyond
a few AU.  Such embryos may have masses comparable to that 
\citep[$M_c \sim 10 M_\oplus$,][]{Pollack.etal.1996} required
to initiate efficient accretion of gas onto the progenitor cores
of proto gas giant planets.  

The actual mass range and radial extent of zones 1 and 2 depend
on the magnitude of $\dot M$.  During the depletion of the disk,
zones 1 and 2 cover smaller planetary masses and radial distance
from their host stars.  

\subsection{Sensitive dependence on the disk models}\label{sec:GL07plot}
\begin{figure}
	\includegraphics[width=\columnwidth]{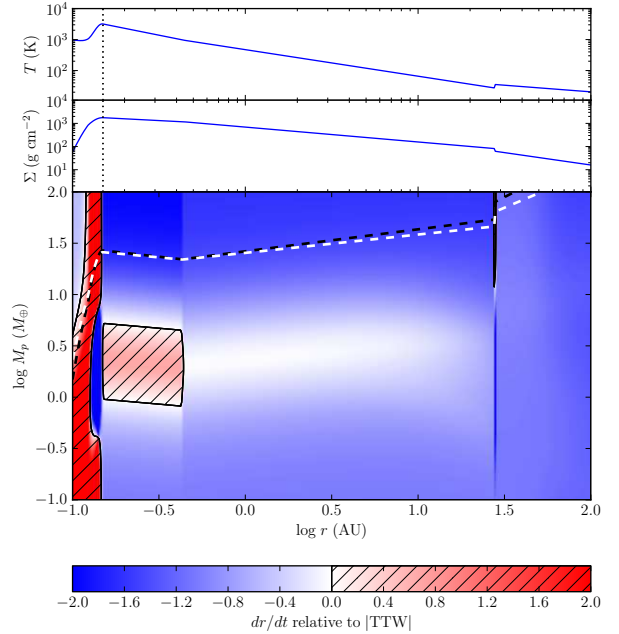}
	\caption{The same as figure \ref{fig:notrack_viscous} for the GL07 model, as described in \S\ref{sec:GL07plot}.}
\label{fig:notrack_GL07}
\end{figure}

In figure \ref{fig:notrack_GL07}, we show the landscape 
for a GL07 disk with $\Mdot=10^{-8} M_\sun \yr^{-1}$, $\alpha=10^{-3}$, 
$\beta=1$.  In the inner most region of the disk viscous heating 
dominates, and it is analogous to zone 1 in the fully viscous 
disk model (see discussions in the previous section).  Consequently,
planets with masses in the appropriate range undergo outward migration.
Outside this region, stellar irradiation provides most of
the heating for the disk and modifies the temperature distribution.
Consequently, the horseshoe torque associated with the co-rotation resonance
is weakened.  There is a distinct strongly-opaque region where the 
horse shoe torques reduce the impact of differential Lindblad resonance 
and significantly slow down but does not reverse planets' inward 
migration, even for those with optimum masses.  The main reason 
for this reduction in the inward migration rate is due to the relatively 
steep decline of the mid plane temperature with the radius.  With 
the inner region being heated by viscous dissipation, the temperature
gradient {\it i.e.}~$q(r)$ in GL07 is more negative than that for a 
totally passive passive, flaring disk in CG97.

For these choices of parameters, the most massive embryos that can be trapped 
is $\sim 3-4 M_\odot$ in an unperturbed disk at a fraction of an AU.  This 
mass range is well below that ($M_c$) required for the acquisition of massive 
gaseous envelope for gas giants through the onset of efficient gas accretion.  
The location is also too close host stars to be compatible with that of most 
known gas giants.  The upper mass limit of retainable embryos increases with
the poorly determined and constrained magnitude of $\alpha$. But, for a given
$\dot M$, the surface density of the gas and presumable the heavy elements
decreases with $\alpha$.  A larger value of $\dot M$ or a smaller value of $\alpha$ may also 
lead to the growth and retention of more massive cores.  The torque applied
by the cores can also lead to modifications in the $\Sigma$ distribution and
this feed-back effect can also reduce the type I migration rate (see \S
\ref{sec:uncertainties}).

\subsection{Disks with dead zones and active layers}\label{sec:constSigmaA}
\begin{figure}
	\includegraphics[width=\columnwidth]{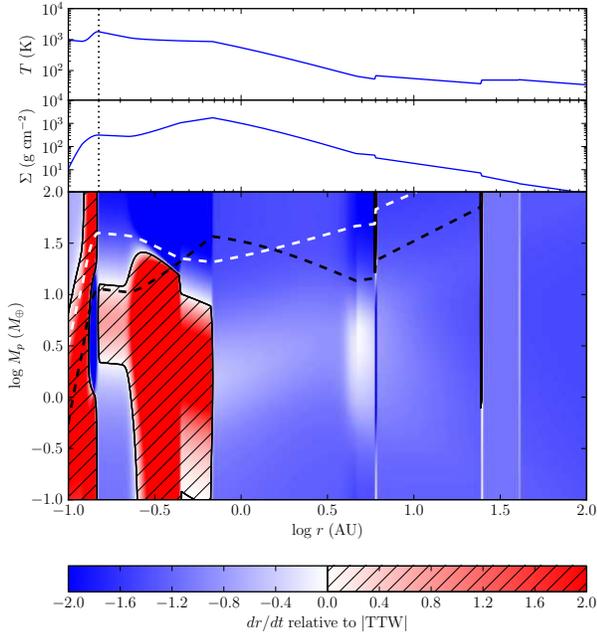}
	\caption{The same as figure \ref{fig:notrack_viscous} for a disk with a dead zone and an active layer with a constant column depth (see \S\ref{sec:constSigmaA}).}
	\label{fig:notrack_deadzone}
\end{figure}

In figure \ref{fig:notrack_deadzone} we show the relative migration 
rates for a planet in a disk with a MRI active surface layer. For 
simplicity, we first consider a model in which the surface density 
$\Sigma_A$ of the MRI active layer is a constant ($\Sigma_A=50 
{\rm g cm^{-2}}$), similar to the seminal \citet{Gammie.1996} paper, 
and potentially there is a dead zone below it with surface density $\Sigma_D$.
We take the viscosity in the active layers to be large with an 
$\alpha_A=10^{-2}$, while the viscosity in the dead zone is small 
$\alpha_D=10^{-4}$.  With this small but non-vanishing value for 
$\alpha_D$, we approximate the propagation of turbulent disturbances
near the disk surface into the midplane regions of the disk.  The 
actual values of $\Sigma_A$ and $\alpha_D$ are still poorly constrained \citep{Fleming.Stone.2003,Turner.etal.2007}.

In regions where $\Sigma_A$ is smaller than the total surface density, 
accretion flow in the active surface layer of the disk leads to a smaller 
effective magnitude for the vertically averaged 
$\alpha\equiv (\alpha_A\Sigma_A+\alpha_D\Sigma_D)/\Sigma$.  
The main consequence of this modification
is the expansion in both the lower and upper mass range for embryos 
that can be retained.  The upper mass range exceeds the magnitude of $M_c$.  
The lower mass range extends well below the isolation mass for embryos 
even at relatively small radius. However, if we take into account the 
effect of vertical structure on the saturation parameter, the lower mass 
range would be limited to a fraction of the Earth mass.  Although these 
tendencies maybe favorable to the formation of gas giants, the location 
of the planet trapping radius is only a fraction of an AU.  Larger values 
$\dot M$ are needed for the planet trapping radius to become comparable 
with the typical orbital semi major axis of known gas giants. 

\section{Some potential complications}\label{sec:uncertainties}

A robust feature in the above results is a tendency for super-Earths to 
migrate outwards in the inner regions and inward in the outer regions 
of the disk.  This property implies the presence of a planet-trapping 
radius where these planets may converge and accumulate.  We also showed 
that the mass range of the potentially trapped planets and the location 
where their type I migration is stalled are model-dependent and subject
to uncertainties.  In this section, we list several additional subtle 
effects which may also affect these quantities.  

\subsection{Disks with variable active layers}\label{sec:varSigmaA}
\begin{figure}
	\includegraphics[width=\columnwidth]{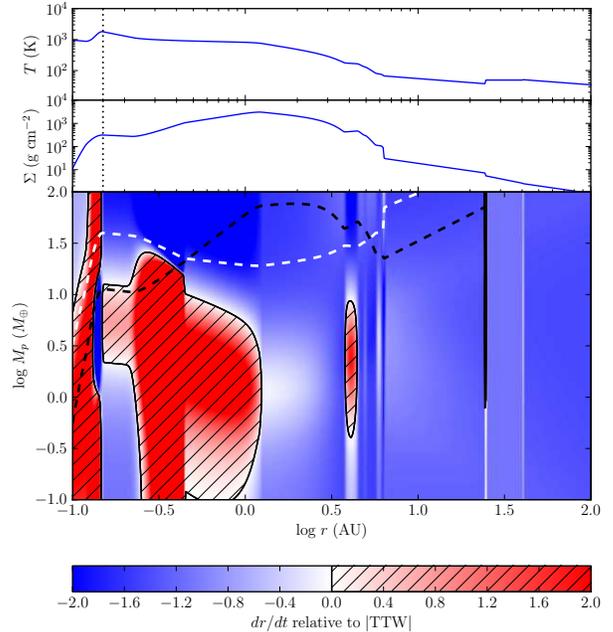}
	\caption{The same as figure \ref{fig:notrack_viscous} for a disk with a dead zone with variable thickness depending on the dust to gas ratio (see \S\ref{sec:varSigmaA}).}
	\label{fig:notrack_snowline}
\end{figure}

In figure \ref{fig:notrack_snowline}, 
we show the relative migration rates for a planet in such a disk with 
a $\Sigma_A$ which varies with the local dust-to-gas ratio.  
We use an approximate fit to the results of \citet{Ilgner.Nelson.2006} 
[model 4] in which
\begin{equation}
	\Sigma_A \simeq \left(\frac{0.0025}{f_d}-0.018\right)
\pfrac{r}{\rm AU}^{2.87},
\end{equation}
where $f_d$ is the dust-to-gas ratio for micron-sized particles.
This prescription introduces a feature at the snow line where the 
dust to gas ratio decreases as a planet moves inwards due to 
the sublimation of ice grains \citep{Kretke.Lin.2007,Kretke.Lin.2010}.  
In this confined transition region,
embryos can be trapped at several AUs from their host stars.  
However, we note that for these parameters the mass of cores 
that can be trapped is relatively small, below $M_c\sim 10 M_\oplus$.
Whether or not these cores are sufficiently massive to accrete 
atmospheres and become gas giants depends upon properties such as 
the planetesimal mass accretion rate and atmospheric 
opacity \citep{Hubickyj.etal.2005,Rafikov.2006}.

\subsection{Revisiting the Horseshoe torques in the Dead Zone}\label{sec:horseshoetorques}
In addition to the reduction in $\alpha$ and the corresponding
modification in the $\Sigma$ and $T$ distribution, a layered disk
structure may also affect the saturation of embedded planets'
co-rotation torque.  The horseshoe region where the co-rotation
torque is applied has a dimensionless width in the disk plane 
$x_s \simeq (M_p r/M_\ast h)^{1/2}$ \citep{Paardekooper.Papaloizou.2009}.
In the direction normal to the disk plane, the horseshoe region 
extend to a similar dimensionless height $z_s \sim x_s$.  

For a 10 $M_\oplus$ planet ($M_p/M_\sun\sim 3\times 10^{-5}$) embedded in
the midplane of a disk with a typical $h/r \sim 0.05$, $z_s \sim 0.02$.  
For these parameters, there is a substantial dead zone that extends 
over one disk scale height (so that the dimensionless height of the 
dead zone $z_d> h/r$) and the entire horseshoe region is engulfed 
within the dead zone.  In this situation, the horseshoe region 
is entirely isolated from the active layer. Ideally, the continuous viscous 
diffusion of disk gas into the horseshoe region of relatively low mass planets
may be quenched if their $z_s < z_d$.  It is unclear if the 
Reynold's stresses which provide a source of angular momentum transport 
in the dead zone (and therefore generate an effective $\alpha_D$) may also 
redistribute vortensity and entropy between the horseshoe region 
and the surrounding disk with a similar efficiency.  Without such 
redistribution, the co-rotation torque of low-mass planets with coplanar 
orbits would be saturated after a few libration periods. 

\begin{figure}
	\includegraphics[width=\columnwidth]{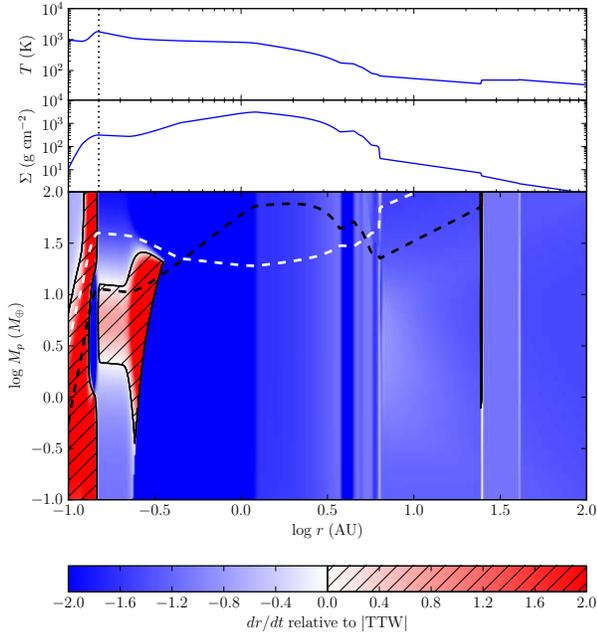}
\caption{The same disk model as shown in figure \ref{fig:notrack_snowline} but with the assumption that the co-rotation torques fully saturate in the dead zone if $z_s<z_d$ (see \S\ref{sec:horseshoetorques}).}
\label{fig:deadzone_novisc}
\end{figure}

In principle, the limited saturation for planets with $h/r> z_s > z_d$
can be estimated from 3D hydrodynamic simulations.  But, such simulations 
are time consuming, especially for such low masses.  Therefore we utilize 
our approximation to highlight how including this effect may impact the 
migration rates. In figure \ref{fig:deadzone_novisc} we demonstrate the 
results if we assume that the effective viscosity is used in determining the 
Reynolds number (${\rm Re}$) in eq.~\ref{eq:p_nu} is negligible if $z_s<z_d$, meaning that, for the sole purpose of calculating ${\rm Re}$, we 
set $\alpha = 10^{-8}$.  Under this assumption, the horseshoe 
torque saturate for nearly the entire dead zone, resulting in 
inwards migration for almost all masses and positions.

\begin{figure}
	\includegraphics[width=\columnwidth]{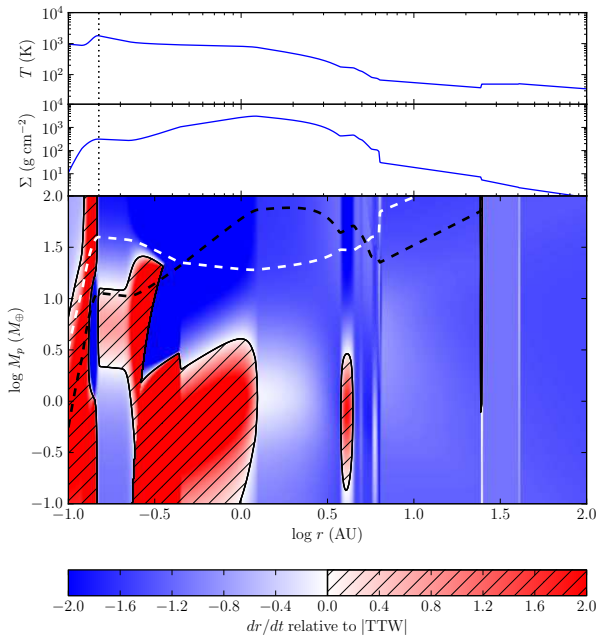}
\caption{The same disk model as shown in figure \ref{fig:notrack_snowline} for the assumption that the effective viscosity to desaturate the corotation torques is equal $\alpha_D$ if $z_s<z_d$ (see \S\ref{sec:horseshoetorques}).}
\label{fig:deadzone_visc}
\end{figure}

As a comparison, in figure \ref{fig:deadzone_visc} we assume that 
if $z_s<z_d$ then the relevant effective viscosity is the dead zone 
viscosity $\alpha_D$.  In this case the horseshoe torques saturate 
at a much smaller mass so we still see stalling locations for small 
planetesimals, but we lose the ability to stall cores greater than 
around 3 $M_\oplus$.

In these estimates, we assume the cores have co-planar orbits.  In 
N-body simulation of planetesimal dynamics and embryos formation,
their mutual gravitational perturbation and collisions can lead
to dispersions in the velocity ($\sigma_z$) normal to the disk plane.  
During the initial growth, $\sigma_z$ of the most massive embryos
is damped by dynamical friction.  But during the advance Oligarchic
growth stage, perturbation between comparable mass embryos may 
induce excitation $\sigma_z$ to be a fraction of their surface escape
speed, despite gas drag \citep{Kokubo.Ida.1998}. If $\sigma_z > \OmK z_d$, 
the embryos would encounter MRI active regions during at least a fraction 
of each Keplerian orbital period and the reduction of the diffusion 
through the horseshoe regions would no longer be an issue.    

\subsection{Feedback and transition to type II migration}
In the above sections, we assume that planets with masses 
up to a few $M_\oplus$ do not significantly alter the average disk $\Sigma$ 
and $T$ distribution. In our evaluation on the direction and rate of 
such planets' type I migration and the extent of saturation ({\it i.e.}~the magnitude and sign of $f$), we calculate the $\Sigma$ and $T$ 
distribution neglecting any feedback effects.  However, we note that 
previous research has 
demonstrated that, even with a modest mass (such as $\sim 5 M_\earth$), 
planets may alter the disk structure, especially in the horseshoe region, 
and slow down the rate of migration \citep{Masset.etal.2006b}.  

Additionally, the type I migration rate is modified once 
the interaction between the disk and a planet become non-linear.  
This limit is reached when the planet's hill sphere becomes 
comparable to the disk scale height, {\it i.e.}, once 
\begin{equation}
\Theta_1 \equiv  \frac{M_p}{M_\ast}\pfrac{r}{h}^3 > 1.
\label{eq:nonlinear}
\end{equation}
Additionally the planet begins to open a gap \citep{Lin.Papaloizou.1993}, 
when
\begin{equation}
	\Theta_2 \equiv  \left(40 \alpha \pfrac{h}{r}^5\right)^{1/2}  
\label{eq:gapopening}
\frac{M_\ast}{M_p} < 1.
\end{equation}
Gap formation interrupts accretion flow through the disk and consequently 
triggers a transition from type I to type II migration \citep{Ward.1997}. This 
limit is more relevant to gas giants rather than super-Earths.  

Though super-Earths do not generally have sufficient mass to open gaps and,  their feedback can 
significantly affect type I migration process. In the low viscosity 
(with $\alpha< <1$) limit, gas in the horseshoe region encounters weak 
shock when it flows close to a-few-Earth-mass planets' Roche lobe.  
Wave dissipation across the shock leads to angular momentum deposition 
and evolution in the $\Sigma$ distribution in the horseshoe 
region \citep{Li.etal.2009}. On the horseshoe libration time scale, 
this effect can significantly reduce the torque magnitude 
and planets' type I migration rate \citep{Yu.etal.2010}.  

Shock induced dissipation along the stream lines also modifies
the vortensity (the vorticity divided by the surface density) along each 
streamline. The development of extrema in the vortensity distribution 
across the stream lines promotes the growth of secondary or Rossby Vortex 
instabilities (RVI) which can be suppressed by adequate viscosity. In the 
limit of negligible viscosity, the growth of RVI can lead to the emergence
of large vortices \citep{Li.etal.2005}.  The torque induced by the 
interaction between the planets and the vortices is chaotic.  It can affect
both the direction and rate of type I migration.  Nevertheless, the onset 
of RVI significantly reduces the effective magnitude of type I migration rate
\citep{Yu.etal.2010}.  These feedback effects can extend the upper mass limit
for the trapped planets.

\section{Relationship to Population Synthesis Models} \label{sec:pop}

In order to assess how these various disk models may impact the emergence and appearance of planetary system one may wish to include these prescriptions in population synthesis models.  We note that without actually preforming a detailed population synthesis model we cannot compare how precisely the models will be affected, but we can compare these models to existing formulations.
In \citet{Ida.Lin.2008a} and \citet{Ida.Lin.2010} the authors assume that the gas follows the radial surface density gradient of the minimum mass solar nebula and the temperature follows a power-law of $T\propto r^{-1/2}$.  As indicated in figure \ref{fig:intro}, this disk structure would lead to inwards migration under both the isothermal and adiabatic regimes.  Therefore if one assumes that disk model, including the additional thermal effect and saturation effects discussed in this paper will not significantly change the result, that type I migration at its predicted values is too fast to form the observed planet population.   
In \citet{Ida.Lin.2008b} where the authors adjust the surface density to include a simple prescription for the dead zone.  However the same thermal profile is still used.  In this case fully considering the possibly of torque saturation causes the snow-line "planet trap" to be only effective for a limited range of masses.
The largest change for these models is that if they used a self-consistent temperature model they could create inner, viscously heated region. In the papers as given the authors recognize that in this region they must artificially stop the planets migration, and the viscously heated inner region may be a natural way to achieve this effect.

In \citet{Mordasini.etal.2011} the authors used the viscous heating prescription of \citet{Lyra.etal.2010} and \citet{Sandor.etal.2011} which is very similar to the model described in \S \ref{sec:model_vis} and figure \ref{fig:notrack_viscous}.  This leads to various ``concentration points'' for planet migration.  However, in this paper we argue that a viscously heated model may not be valid for large regions of the disk.  If indeed the disk is passively heated from the star then the disks will not have planet traps far from the parent star unless there is dead zone or other change in viscosity.  
If this is the case than the results may be more similar to those shown in \citet{Mordasini.etal.2012} in which the authors use a viscously heated model but do not account for the thermal effects.  This means that they see inwards migration throughout the disk without stalling zones. 

\section{Summary and Discussion}\label{sec:summary}
In this paper we discuss the migration of low-mass planets (super-Earths) 
in gaseous disks via type I migration including the impact of horseshoe 
torques as presented by \citet{Paardekooper.etal.2010a}.
We find that planets with a few $M_\oplus$'s have a tendency to 
undergo outward migration through regions which are predominantly
heated by viscous dissipation and inward through regions which are
mainly heated by stellar irradiation.  Since typical protostellar 
disks are mostly heated by viscous dissipation at small radii and 
by stellar irradiation at large radii, these planets tend to converge
and stall at a fraction of 1 AU from their host stars.  This 
result is robust and general for self-consistent protostellar disk 
structures in accordance with the conventional $\alpha$ models.  
Thus, we anticipate a high retention efficiency for super-Earths.
This extrapolation is consistent with the observed ubiquity of 
super-Earths.  

However, the location of the stalling radius is somewhat model-dependent.  
Our numerical results indicate that its value is smaller in 
models that take into consideration stellar irradiation than in models that are primarily heated by viscous dissipation \citep[e.g.][]{Lyra.etal.2010}.  
In the composite (viscous dissipation plus surface irradiation) models,
it is difficult to sustain outward migration to large radii without 
any self-shadowed effects.  While self-shadowing by a puffed up inner 
rim is possible \citep{Ke.etal.2011}, it is unclear if this shadowing can extend beyond a few AUs. 

In all of the models the planets migrate inwards or outwards until 
they reach a barrier such as inner-edge of the dead zone $r_{id}$, the 
magnetospheric truncation radius, or other stalling radii.  
However, even if a planet stalls at a given radius within the 
disk it may not determine its final location because, as a disk 
evolves, the location of these barriers tend to change.
For example, $\rmag$ increases and $r_{id}$ decreases with 
diminishing $\dot M$. 
This sensitive dependence on the disk structure
may lead to large dispersion in the asymptotic period of super-Earths.
In contrast to a 3-day pile up in hot Jupiters' period distribution,
super-Earth candidates, which are identified by Kepler's mission, 
show a nearly uniform logarithmic period distribution without any
preferred orbital configuration.  The observed spread-out period 
distribution of super-Earths does not contradict the expected large 
dispersion in both the disk structure and planets' stalling condition.

For planets that do reach the inner-edge of the disk, the stalling 
locations for low-mass planets differs from those of more 
massive (Jovian mass) gap-opening planets. These high-mass planets 
terminate their type II migrating once they 
have reached a 2:1 resonance interior to the inner edge of the 
truncated disk or in the limit that the disk mass 
becomes smaller than the planet 
mass \citep{Lin.etal.1996}.  Therefore we expect that short period 
high mass planets (hot Jupiters), the product of type II migration, 
should be found 
on average interior lower mass planets, whose migration is stalled 
at the magnetospheric truncation radius.
As we continue to detect low-mass planets on short-period orbits 
they will become a important tool to constrain models of planet 
formation and migration.
The final stopping location of short-period planets will reveal 
information about the structure of the inner disk region.

Additionally, as type I migration is proportional to the planetary 
mass, planets on the upper range 
of the type I migration regime are the most vulnerable.
For these higher-mass embryos the horseshoe torque begin to saturate, 
meaning that they migrate inwards at the full rate.  However, these 
embryos are also the planets which are able to modify the 
$\Sigma$ profiles of the gas disk in their vicinities.  These feedback
effect could quench type I migration, induce vortex formation through
RVI,  or lead to type II migration.  These effect may lead to the 
fall off in the observed size distribution (at around 3-4 Earth radii).

High mass-accretion rate promotes core retention at large radii.  
This correlation is primarily due to the fact that the aspect ratio 
($h/r$) generally increases with $\dot M$.  In this limit, the horseshoe 
torque does not saturate until the embryos have acquired a relatively
large planet mass. For sufficiently large $\dot M (> 10^{-7} M_\odot$ 
yr$^{-1}$), stalled embryos can attained sufficiently large mass 
to initiate efficient gas accretion.  A larger $\alpha$ can also 
prevent inwards migration of these particularly vulnerable cores.  
However, the location of the transition from the viscously heated 
region to the strongly opaque region (i.e.~the transition from 
inwards to outwards migration) depends only on $\Mdot$ not on $\alpha$.
In general the existence of a dead zone hinders the formation of cores 
at large radii because the low viscosity prevents the barrier for high-mass
cores.  However, the additional degree of freedom provided 
by the variable $\alpha$ parameter open the possibility of rapid outward migration for embryos with various masses.
This effect may create planet traps, but only if the horseshoe torques can remain unsaturated.

A model in which planetary cores are retained at the ``planet traps'' begs 
one very important question.  
Once a planetary embryo acquires more than one Earth mass, it would migrate to one of these trapping locations on relatively short timescales, unless there is another mechanism to slow type I 
migration.  Therefore, if, as expected, 
a system of multiple embryos emerge in a 
disk around a common host star, they would have a tendency to migrate to
locations which are already occupied by others. This convergent dynamical 
evolution raises the question of how two (or more) embryos interact as 
they approach one another. 
This interaction needs to be explored in future studies, especially as it is 
clear that super-Earths appear to primarily be found in multi-planet systems.

The prescription generated in this paper can be applied to 
population synthesis models\citep{Ida.Lin.2004,Ida.Lin.2010}. 
Through statistical comparisons between
these models and the observed data, it is possible to highlight both
robust physical processes which lead to common features and marginal 
effects which introduce diversity and dispersion.  

We thank Clement Baruteau for useful conversation.
This work is supported in part by NASA grants NNX07AI88G, NNX08AL41G 
and NNX08AM84G as well as a NSF grant AST-0908807.
K.~K.~appreciates funding from the NASA OSS program (P.I. H.~Levison).


\appendix
\section{Generalization of GL07}\label{sec:appendix}
While GL07 is presented for a disk model with a constant $\alpha$ parameter, it can be generalized to include radial dependence.
The temperature models presented in GL07 are entirely local, and therefore independent of the radial variation of $\alpha$.
So for an arbitrary surface density profile one may simply use the prescriptions presented in GL07.
However, in order to determine the surface density profile, assuming a steady-state disk, 
\begin{equation}
	p=-q -\zeta - \frac{3}{2}
\end{equation}
We define $\zeta \equiv d\ln\alpha/\ln r$.
In the optically thin, marginally opaque, and weakly opaque regions the temperature profile is independent of the surface density profile so one can directly compute the surface density profile from this relationship.
In the strongly opaque and viscously heated region the temperature gradient depends on the surface density
\begin{equation}
	p_{\rm SO} = -\frac{15n+15-9\beta + \zeta(14n+14-2\beta)}{2(7n+15-\beta)}
\end{equation}
where $n$ is the polytropic index and the opacity varies according to $\kappa(T) = \kappa_V (T/T_*)^\beta$.
\begin{equation}
	q_V = \frac{9n-3\beta+15-4\zeta}{4n+8}-3
\end{equation}
For the fiducial model in which $n=2$ and $\beta=1$, 
\begin{equation}
p_{\rm SO}  = -\frac{9+10\zeta}{14},
\end{equation}
and 
\begin{equation}
q_{\rm SO}  =-\frac{6+2\zeta}{7}.
\end{equation}
In the viscously heated region
\begin{equation}
	p_{\rm V} = -\frac{3+6\zeta}{8}
\end{equation}
and
\begin{equation}
	q_{\rm V} = -\frac{9+2\zeta}{8}.
\end{equation}

We note that these are valid only under the same limit of GL07 that 
\begin{equation}
	\frac{h}{r} \frac{d \ln (Z \rho_m)}{d \ln r} \ll \frac{z_s^2}{h^2}r\frac{d}{dr}\pfrac{h}{r}
\end{equation}

\end{document}